\documentstyle{article}
\begin{document}
\begin{center}
{\huge Primordial Black Holes With  Variable Gravity}
\end{center}
\vspace{2cm}
\begin{center}
{\Large Arbab I. Arbab}\footnote{E-mail:  arbab64@hotmail.com}
\vspace{1cm}\\
Department of Physics, Faculty of Science, University of Khartoum,
P.O. Box 321, Khartoum 11115, SUDAN\\
\end{center}
\vspace{4cm}
\centerline{ABSTRACT}
\vspace{0.5cm}
We have studied the evolution of primordial black holes (PBHs) in a universe with
a variable gravitational constant and bulk viscosity. 
We have found that the strength of gravity has changed appreciably in 
the early universe. The gravitational constant attained its greatest value at 
$t=10^{-23}$sec after the Big Bang.
PBHs formed at the GUT and electroweak epochs would have masses about 
$\rm 1.85\times 10^9 g$ and $\rm 1.8\times 10^{-7}\ g$ respectively. 
Their temperatures when they explode are $6\times 10^8\rm\ K$ and $\rm 6 K$
respectively.
PBHs formed during nuclear epoch are hard to detect at the present time.
The gravitational constant ($G$) is found to increase as $G\propto t^2$ in 
the radiation epoch. The gamma rays bursts (GRBs) may have their  origin
in the evaporation of the PBHs formed during GUT time.
\\

\vspace{2cm}
PACS no(s): 98.80.Dr., 98.80.Hw, 98.80.-k
\newpage
\large
{\bf 1. INTRODUCTION}
\vspace{0.5cm}

As well as being formed in the  course of natural stellar or galactic evolution,
black holes may also have been produced primordially, i.e., at the very earliest
epochs of cosmological time. The process of a primordial black hole (PBH) formation
in a cosmological model with variable gravitational constant $G$ creates an
interesting problem.
If one considers a Schwartzschild's black hole formed in the very early Universe
at  time $\rm t_f$ when $G$ had the value $\rm G_f$, which may be different from the
present gravitational constant $G_0$. The horizon size of a black hole is given
by $\rm R_f=\frac{2G_fM}{c^2}\sim ct_f$, where $M$ is the mass of the black hole.
The horizon area $(A$) and the entropy $(S)$ of the black hole are
given by
\begin{equation}
A\propto M^2G^2 \  \ \rm and \ \ S\propto GM^2\ .
\end{equation}
Thus as long as $G$ is increasing the entropy of the black hole increases.
However, in the case of a decreasing $G$ one needs a further remedy 
to account for an increased entropy.
Barrow [1] has conjectured the idea of a gravitational memory of the black hole
(i.e., a black hole remembers the value of $G$ at the time of its formation).
This will have a dramatic implications for the Hawking evaporation of
PBHs as the temperature and the life time of the evaporating black holes are
determined by the value of $\rm G_f$ and not by $G_0$. These are given by [2]
\begin{equation}
\rm\tau\sim G_f^2M^3 \ \ \rm and\ \ T\sim G_f^{-1}M^{-1}\ .
\end{equation}
Black holes which explode today are those whose Hawking's life time is equal to
$t_0$ (the present age of the universe).
Consequently one obtains
\begin{equation}
\rm M_{\rm ex}\simeq 4\times 10^{14}(\frac{G_0}{G_f})^{\frac{2}{3}}\ \rm g\ ,
\end{equation}
and their temperature when they explode is therefore given by,
\begin{equation}
\rm T_{\rm ex}\simeq24\times(\frac{G_0}{G_f})^{\frac{1}{3}}\ \ \rm MeV\ ,
\end{equation}
where the suffix ``ex" stands for explode.
\\

\vspace{0.5cm}
{\bf 3. EVAPORATION OF PRIMORDIAL BLACK HOLES}

\vspace{0.5cm}

The radiated particles from a PBH have temperature $T$ given by [2]
\begin{equation}
kT=\frac{\hbar c^3}{8\pi GM}\sim 10^{26}M_g^{-1}
\end{equation}
and the entire mass is radiated away in a time given by
\begin{equation}
\tau\sim 3\times 10^{-27}M_g^3\ \rm sec,
\end{equation}
where $M_g$ is the mass in grams.
It is clear that as $M$ decreases $T$ increases and the mass
loss increases until finally reaches a catastrophic limit
(explosion or evaporation).
Note that the stellar black hole $(M_g>10^{23}$) is unlikely to
explode in the life time of the Universe [3].
Carr [4] has investigated the PBH formation and evaporation in order to see
whether the presently observed nucleon density as well as the microwave
background radiation (MBR) can be explained in terms of emission of
baryons, leptons, photons and so on, by a low mass black hole. PBHs formed
from inhomogeneities at time $t$ must have an initial mass ($M_i$)
of the order of the particle horizon mass ($M_H$) [5]:
\begin{equation}
M_i\approx M_H=c^3 G^{-1}t=10^5(t/s)M_\odot
\end{equation}
where $M_\odot$ is the solar mass.
PBHs forming at Planck time ($10^{-43}$ sec) would have the
Planck's mass ($10^{-5}$g), whereas those formed at $10^{-23}$ sec would
have a mass $10^{15}\rm g$ required for PBHs which evaporate at the present
epoch. 
The size of the PBH at any given time is limited by the size of the particle
horizon. PBHs can radiate either elementary particles, e.g., quarks,
gluons, which later emit  particles such as baryon, meson and leptons;
or composite particles directly (baryon, mesons, leptons).
A PBH does not matter whether the emitted  particle is a particle or an
anti-particle. So the baryon number is not necessarily conserved.
This possibility can be used to account for the observed baryon-to-photon
ratio.
\\

\vspace{0.5cm}
{\bf 2. MACH'S PRINCIPLE AND THE VARIABLE GRAVITY}\\
\vspace{0.5cm}

The inertial forces observed locally in an accelerated laboratory may be
interpreted as gravitational effects having their origin in distant matter
accelerated relative to the laboratory [6,7].
Einstein has tried to incorporate this principle in the formulation of his
theory of general relativity (GR). Brans and Dicke have developed a theory
which incorporates Mach's principle. A model incorporating the elements
of Mach's principle was given by Sciama [8]. He, from dimensional 
argument, concluded that the gravitational constant $G$ is related to the
mass distribution in a uniform expanding Universe through
\begin{equation}
\frac{GM}{Rc^2}\sim 1
\end{equation}
where $R$ and $M$ are the radius and the mass of the visible Universe,
respectively.
This relation suggests that either the ratio $M$ to $R$ should be fixed or
the gravitational constant $G$ observed locally should be variable and determined
by the mass distribution. Only mass ratio can be compared at different points, but
not masses.
It should be stated that the strong equivalence principle upon which GR stands
is incompatible with variable $G$.
In 1937 Dirac postulated the existence of very large numbers and
constructed a cosmological model in which $G$  decreases with time as
$G\propto t^{-1}$ in order not to change the atomic physics [12].
Unfortunately, his model could not resist the observational data.

We have recently presented a cosmological model with variable $G$ and bulk
viscosity. The gravitational constant  $G$ is found to vary with time as [9],
\begin{equation}
G\propto t^{\frac{2n-1}{1-n}}\ , 
\end{equation}
where $n$ is the viscosity `index', $0\le n\le 1$, and $G\propto \exp(Bt)$
, where $B=\rm const$., during inflation ($n=1)$.

\vspace{1cm}
{\bf 3. STRONG GRAVITY}
\vspace{0.5cm}

Salam [10,11] has considered the gravitational interaction mediated via heavy
mesons and found that the gravitational forces are very strong. He remarked that
the nuclear physics should better be called {\it strong gravity}.
Sivaram and Sinha [11] identified the strong f-gravity metric with Dirac's
atomic metric and the large value of the coupling constant (that is $G_f=10^{40}G_0$)
provided the physical basis for the Large Number Hypothesis (LNH).
So if one considers the earliest era when the Universe consisted of
an extremely hot compact gas of hadrons, the epoch $10^{-23}$ sec, then
if $G$ varied according to LNH right down to the present epoch, it would have a very
large value $G=10^{40}G_0$, at the beginning of the hadron era. This value is
precisely the value $G_f$ found in considering the short range f-gravity
mediated by massive $2^{+}-f$ mesons. Thus in a region of strong
curvature, nuclear physics is analogous to gravity.\\
It is evident from eq.(8) that if one considers the Planck epoch, the nuclear
epoch and the present epoch we will get
\begin{equation}
\rm\frac{G_N}{G_{Pl}}\frac{M_N}{M_{Pl}}=\frac{R_N}{R_{Pl}}
\end{equation}
and
\begin{equation}
\rm\frac{G_0}{G_{Pl}}\frac{M_0}{M_{Pl}}=\frac{R_0}{R_{Pl}}
\end{equation}
where the $\rm G_N (G_{\rm Pl}), M_N (M_{\rm Pl}), R_N (R_{\rm Pl}) $ are the values
 of the gravitational constant, mass and the radius of the Universe at the
 nuclear (Planck) epoch, respectively.
It has been shown that throughout all epochs in the early universe, the relation [7]
\begin{equation}
Gm^2=\hbar c=\rm const.
\end{equation}
was valid, i.e., we had
\begin{equation}
\rm G_N m_N^2=G_{\rm Pl}m^2_{\rm Pl}=G_Wm_W^2=G_{\rm GUT}m^2_{\rm GUT}=\hbar c \ ,
\end{equation}
where $\rm G_N, G_W, G_{GUT}$ refer to the strong (nuclear) gravitational, weak and GUT
coupling constants respectively and $\rm m_p, m_{Pl}, m_W$ and $\rm m_{GUT}$
refer to the nucleon mass, the Planck's mass, the intermediate boson and GUT 
unification mass respectively.
Inserting the numerical values: $R_0=10^{28}$cm, $R_N=10^{-13}$cm,
$\rm R_{Pl}=10^{-33}$cm, $M_0=10^{56}$g, $m_{\rm Pl}=10^{-5}$g, 
 $\rm m_{W}=10^3\ GeV$ and $m_{\rm GUT}=10^{15}\rm GeV$: 
eqs.(10), (11) and (13) yield
\begin{equation}
\rm G_{\rm GUT}=10^{8}G_0\ ,\ G_{\rm W}=10^{32}G_0\ , \ G_{\rm N}=10^{40}G_0, \ \ \ \rm and\ \ \ G_{Pl}=G_0\ .
\end{equation}
This behavior of $G$ can not be interpreted by Dirac or Brans-Dicke model [12].\\
Barrow and Carr [5] have considered the evolution of the PBHs in the context of
scalar-tensor theories and in particular to Brans-Dicke theory. The cosmological
considerations investigated by them restrict the value of $\omega$ (the coupling
constant) to the value $-4/3 >\omega >-3/2 $. The case $\omega=-4/3$ was,
however, excluded. Comparing these results with our model [9], we obtain
the same constraint for $\omega $ in addition to the physical significance of the
case $\omega=-4/3$. This case corresponds, in our model, to the
inflationary solution. In fact, our temporal behavior of $G$ and the scale
factor $R$ is defined for all values of $n$ (the viscosity index).
The constraint made by Barrow and Carr on $\omega$ would imply an increasing
gravitational constant.\\
 From eq.(9), it can be seen that $G$ increases for  $n>1/2$, decreases
for $n<1/2$,  remains constant for $n=1/2$, and increases exponentially
during inflation, i.e.,  when $n=1$ [9]. 
Our model predicts that in the radiation epoch, for $n=\frac{3}{4}$, 
$R\propto t, T\propto R^{-1}, G\propto t^2$ and $\rho\propto t^{-4}$. 
A similar variation is found by Abdel Rahman [13] in the radiation epoch. 
Thus for Planck, GUT, nuclear and electroweak epochs, one has
\begin{equation}
\rm G_N/G_{Pl}=(t_N/t_P)^2=10^{40}\ , \
\rm G_N/G_W=(t_N/t_W)^2=10^8\ , \
\rm G_N/G_{GUT}=(t_N/t_{GUT})^2=10^{32}\ 
\end{equation}
 and
\begin{equation}
\rm R_N/R_{Pl}=t_N/t_{Pl}\ , \ R_N/R_W=t_N/t_W\ , \ R_N/R_{GUT}=t_N/t_{GUT}\ ,
\end{equation}
or
\begin{equation}
\rm R_N=10^{-13}\ cm\ , R_W=10^{-17}\ cm\ , R_{GUT}=10^{-29}\ cm\ ,\ R_{Pl}=10^{-33}\ cm\ ,
\end{equation}
where $\rm t_N=10^{-23} sec, t_P=10^{-43} sec, t_W=10^{-27} sec$ and 
$\rm t_{GUT}=10^{-39} sec$, are the nuclear time, Planck's time, electroweak 
time and GUT  time. Thus both Abdel Rahman's model and the present model 
predict the behavior of  $G$ quoted in eq.(14).
The relation $G\propto t^2$ is equivalent to eqs.(8) and (12).
Note that in the Standard Model $R\propto t^{1/2}, T\propto R^{-1}, 
G=\rm const.$ This relation can't account for the above relations.
Thus it is suggestive to use this equation to predict a fundamental mass 
at any epoch in the early universe.
If one combines this law with eqs.(8) and (12), one finds
\begin{equation}
m\propto t^{-1}\ \ , 
\end{equation}
a relation that was valid in the early universe.
It has been suggested by several authors that the PBH formed at the
nuclear epoch ($10^{-23}$ sec) would have a mass of $10^{15}$g.
However, if one considers the `correct' value of $G$ we would obtain a value
of $10^{-24}$g. This is, in fact, equal to the mass of the proton.
PBHs formed in the early universe would have temperatures
and masses, when exploded, given by (eqs.(3), (4) and (14))
\begin{equation}
\rm T^N_{\rm ex}=1.29\times 10^{-2}\ K\ \  and \ \ M^N_{\rm ex}=8.6\times 10^{-13}\ g\ ,
\end{equation}
\begin{equation}
T^{\rm GUT}_{\rm ex}=6\times 10^8\ \rm K\ \ and \ \rm M^{\rm GUT}_{\rm ex}=1.85\times 10^9 g\ , 
\end{equation}
and
\begin{equation}
\rm T^W_{ex}=6\ \ K \ \ and  \ \  M^W_{ex}=1.8\times10^{-7} g \ .
\end{equation}
One would therefore expect to observe PBHs emitting x-rays. These PBHs
were formed during the GUT phase transition. 
The remnant of PBHs that formed during the nuclear epoch would be difficult 
to detect since they have a temperature below the cosmic background radiation. 
It is interesting to note that PBHs forming during Planck's epoch are not affected, 
since $\rm G_{Pl}=G_0$. Recently, Cline [14] has considered the PBH evaporation 
during the quark-gluon phase transition. He has shown that short gamma rays 
burst (GRB) occurs when the mass of the PBH is either $10^{14}$ or $10^9$ g. 
Thus eq.(20) may indicate the emission of short GRBs.
Hence, the spectra of these PBHs are different from those with constant
$G$. We would, therefore, expect to observe PBHs at a lower temperature.
Thus a possible variation of $G$ would alter the picture of the PBHs previously known.\\

\vspace{1cm}
{\bf ACKNOWLEDGMENTS}
\\

I would like to thank the Abdus Salam International Center for Theoretical Physics
for hospitality and the Associate Scheme for financial support.
\\
\vspace{1cm}

{\bf REFERENCES}
\vspace{0.5cm}\\
1- J.D.Barrow, {\it gr-qc /9711084}  \\
2-  S.W. Hawking, {\it Nature 248}(1974) 30 \\ 
3-  G.Kang, {\it Phy. Rev.D54}(1996) 7483 \\ 
4-  B.J.Carr, {\it Astrophy. Journal 201}(1975) 1\\
5-  B.J. Carr and S.W.Hawking, {\it Mon.Not. Roy. Astr. Soc,168} (1974) 399,
\\ J.D. Barrow and B.J.Carr, {\it Phys. Rev.D54}(1996) 3920,  J.H. MacGibbon, {\it Nature 329}(1987) 308  \\ 
6- T.Singh and L.N. Rai, {\it Gen. Rel. Gravit.15} (1983) 875  \\
7-  Venzo De Sabbata, {\it Acta Cosmologica-Z.9}.(1980) 63 \\ 
V. de Sabbata and C. Sivaram, {\it Astrophys. Spc. Sci.158} (1989) 347\\
8-  D.W.Sciama, {\it Mon. Not. Roy. Astr. Soc. 113} (1953) 34  \\ 
9- A.I.Arbab, {\it Gen. Rel.. Gravit.29} (1997) 61  \\
10- A. Salam, J. Strathdee,{\it  Lett.Nouvo. Cimento. 4} (1970) 101  \\ 
11- C. Sivaram and K.P.Sinha   {\it Phy. Lett.60B} (1976) 181\\
12- P.A.M. Dirac, {\it Nature 139} (1937) 323, C.Brans and R.H.Dicke, {\it Phys.
Rev.124} (1961) 925 \\
13- A.-M.M. Abdel Rahman, {\it Gen. Rel. Gravit.22} (1990) 655\\
14- D.B. Cline, {\it Nuclear Physics A610} (1996) 500c
\end{document}